\title{\normalfont A Note on the Kinetics of Diffusion-mediated Reactions}
\author{
        {\em K. Razi Naqvi\/}\\
                Department of Physics,
        Norwegian University of Science and Technology\\
        N-7491 Trondheim, Norway
         }
\date{\today}

\documentclass[11pt,reqno]{article}
\usepackage{amsmath,amssymb,amsfonts} 
\usepackage{amssymb}  
\usepackage{makeidx}  
\usepackage{graphicx} 
\usepackage{epsfig}
\usepackage[sort&compress,numbers]{natbib}
\usepackage{pstricks}
\usepackage{soul}  
\usepackage[dvips]{curves}  
\usepackage{xcolor}
\usepackage{listings}
\usepackage{amsbsy}

\usepackage{multicol}
\usepackage{pstricks, pst-text}                  
\usepackage{color}
\usepackage{fancyhdr}
\usepackage{graphicx}
\newcommand{\linkcolor}{blue}
\usepackage[colorlinks=true,linkcolor=\linkcolor,urlcolor=\linkcolor]{hyperref} 

\def\d{{\,\rm d}}

\graphicspath{%
    {converted_graphics/}
    {/}
    {Graphics/Chap13/}
    {Graphics/}
}

\begin{document}

\maketitle

%
\vspace*{7ex}

\noindent The prevalent scheme of a diffusion-mediated bimolecular reaction $A+B\rightarrow P$ is an adaptation of that proposed by Briggs and Haldane for enzyme action [{\em Biochem J.\/}, 19:338--339, 1925]. It assumes that a molecule of $A$ and one of $B$ form an encounter complex, $\{AB\}$, which disappears through two competitive first-order reactions, and subjects [${\{AB\}}$], the concentration of the complex, to the steady-state approximation, which leads to the following expression for the rate constant: $k=k_{\mbox{\scriptsize e}}k_{\mbox{\scriptsize r}}/(k_{\mbox{\scriptsize r}}+k_{\mbox{\scriptsize b}})$, where $k_{\mbox{\scriptsize e}}$, $k_{\mbox{\scriptsize r}}$ and $k_{\mbox{\scriptsize b}}$
are the rate constants for the reactions $A+B\rightarrow \{AB\}$, $\{AB\}\rightarrow P$, and $\{AB\}\rightarrow A+B$, respectively.
Since the above expression for the rate constant can be rephrased as $k=\alpha k_{\mbox{\scriptsize e}}$, with $\alpha=k_{\mbox{\scriptsize r}}/(k_{\mbox{\scriptsize r}}+k_{\mbox{\scriptsize b}})$, the foregoing treatment encourages the view that $\alpha$ can be interpreted as the reaction probability per encounter.
 
The purpose of this Note is to explain, {\em by using an argument involving no mathematics\/}, why the breakup of the encounter complex cannot be described, except in special circumstances, in terms of a first-order process $\{AB\}\rightarrow A+B$. Briefly, such a description neglects the occurrence of re-encounters, which lie at the heart of Noyes's theory of diffusion-mediated reactions. The relation $k=\alpha k_{\mbox{\scriptsize e}}$ becomes valid only when $\alpha$ (the reaction probability per encounter) is very much smaller than unity (activation-controlled reactions), or when $\beta$ (the re-encounter probability) is negligible (as happens in a gas-phase reaction). References to some works (by the author and his collaborators) which propound the correct approach for finding $k$ are also supplied.  \\[2ex]

{\large
\begin{multicols}{2}
Consider a liquid solution containing two species, $A$ and $B$, which are capable of reacting with each other to form a product. We may represent this reaction by the following scheme:\\[-1ex]
\begin{equation}\label{eq:BimolReac}
A+B\xrightarrow []{\hspace{1ex}\displaystyle{k}\hspace{1ex}}P,
\end{equation}
which means that the rate of formation of the product may be written as\\[-1ex]
\begin{equation}\label{eq:RateBMR}
\frac{\d P}{\d t}= k[A][B].
\end{equation}
If we want to think about the reaction in detail, we may envisage the course of events as follows: On account of random displacements, solute molecules diffuse through the solvent, and there will arise situations when two solute molecules, one of each type, would become contiguous, surrounded by a shell or {\em cage\/} of solvent molecules. The symbol $X$ ($X=A, B, P \ldots $) will henceforth denote both a reactant/product and a molecule of this reactant/product. We will say that an {\em encounter\/} takes place, or an {\em encounter pair\/} is formed, when two molecules become trapped in the same solvent cage; an encounter is the analogue of a bimolecular collision in the gas phase. Two molecules $A$ and $B$ will be said to be encounter partners when they are in the same solvent cage, and this situation will be represented by the symbol $\{AB\}$ and the formation of $\{AB\}$ will be called the formation of an {\em encounter pair\/}; $A$ and $B$ undergo a number of collisions during an encounter. We owe this evocative account of solute dynamics to Rabinowitch and Wood \cite{Rabinowitch1936TFS,Rabinowitch1937TFS}. Now, the same process of diffusion, which brought $A$ and $B$ into a single  cage, will also tear them apart, and one must ask what happens when two molecules separate from a non-reactive encounter.

If two molecules in the gas phase collide and separate without reacting, the probability that they will ever react is negligibly small because a re-collision between the same molecules is a highly unlikely event. It will be expedient to apply the term {\em encounter\/} also to a bimolecular collision in the gas phase, and to notice that re-encounters, which occur with negligible probability in the gas phase, occur with a high probability in the liquid phase. If the two molecules in a liquid medium do not react during the lifetime of the encounter pair, they will separate, but they may become encounter partners once more, and so on, until the reaction occurs while they are encounter partners, or the two drift so far from each other as to become part of the bulk solution again.

Consider an extreme limit in which the rate constant $k_{\mbox{\scriptsize r}}$ for the reaction 
\begin{equation}\label{eq:EncoPair}
\{AB\} \xrightarrow [\hspace{6ex}]{{\displaystyle k_{\mbox{\scriptsize r}}}} P
\end{equation}
is so large that the reaction occurs before the break up of the encounter pair. In this case, we can replace the two-step reaction
\begin{equation}\label{eq:TwoStep}
A+B\xrightarrow [\hspace{6ex}]{{\displaystyle k_{\mbox{\scriptsize e}}}}\{AB\} \xrightarrow [\hspace{6ex}]{{\displaystyle k_{\mbox{\scriptsize r}}}} P,
\end{equation}
by that shown below ($k_{\mbox{\scriptsize e}}$ denotes the rate constant for the formation of encounter pairs):
\begin{equation}\label{eq:OneStep}
A+B\xrightarrow [\hspace{6ex}]{{\displaystyle k_{\mbox{\scriptsize e}}}} P.
\end{equation}

We now go on to consider the situation where the breakup of the encounter pair can compete with reaction \ref{eq:EncoPair}. The standard procedure is to express the situation under consideration by the following scheme:
\begin{equation}\label{eq:ThreeArrows}
A+B\hspace{1ex}
 \raisebox{0.4ex}{$\xrightarrow [\hspace{6ex}]{{\displaystyle k_{\mbox{\scriptsize e}}}}$}\hspace{-7.2ex}
\raisebox{-0.4ex}{$\xleftarrow [{\displaystyle k_{\mbox{\scriptsize b}\;}}]{\hspace{6ex}}$}
\hspace{1ex}\{AB\}
\xrightarrow [\hspace{6ex}]{{\displaystyle k_{\mbox{\scriptsize r}}}} P
\end{equation}

It is easy to see that, if one imposes the steady-state approximation, scheme \ref{eq:ThreeArrows} leads to the following value of the rate constant \cite{Truhlar1985JCE}:
\begin{equation}\label{eq:wrongrate}
k=\frac{ k_{\mbox{\scriptsize r}} }{k_{\mbox{\scriptsize r}}+k_{\mbox{\scriptsize b}}}k_{\mbox{\scriptsize e}}=\alpha k_{\mbox{\scriptsize e}},
\end{equation}
where $\alpha=k_{\mbox{\scriptsize r}}/(k_{\mbox{\scriptsize r}}+k_{\mbox{\scriptsize b}})$ is the probability that the reaction would occur during an encounter. 

Though Eq.~\ref{eq:wrongrate} appears intuitively obvious and reasonable, the conclusion that the rate constant $k$ equals the product of the reaction probability per encounter and the rate constant for encounter formation is not valid in general (see below), and contradicts careful analyses of diffusion-mediated reactions \cite{KRC1979CPLwidespread,KRN1980JPC,KRN1982PRL,KRN1982JPC,KRN1982CPL1,KRN1982CPL2,KRN1983JCP,KRN1984Colloid}.

In a 1949 article \cite{Collins1949Kimball} Collins and Kimball (C\&K) pointed out that ``when not every collision [read {\em encounter\/}] between particles is effective, the Sveshnikoff procedure of multiplying the flux [a quantity proportional to $k$] by $\alpha$ is not self-consistent", but this statement has evidently been overlooked by most of those who cite their article. I would like to add here, for the sake of historical veracity, that what C\&K call the ``Sveshnikoff procedure" can in fact be traced to Smoluchowski himself \cite{Smoluchowski1917ZPC,Smoluchowski1917KZ}, who used the symbol $\epsilon$ instead of $\alpha$.
 
That there is a fatal flaw in scheme \ref{eq:ThreeArrows} can be established without any mathematical manipulations. One only has to notice that when one writes the first step of this reaction as
\begin{equation}\label{eq:Formation}
A+B\hspace{1ex}
\xrightarrow [\hspace{6ex}]{{\displaystyle k_{\mbox{\scriptsize e}}}}\hspace{1ex}\{AB\},
\end{equation}
it is understood that $A+B$ stands for the situation where the reactant molecules, namely $A$ and $B$, are {\em randomly distributed in the bulk of the solution\/}. It is a mistake, therefore, to represent the breakup of the encounter complex by the reaction
\begin{equation}\label{eq:Breakup}
A+B\hspace{1ex}
\xleftarrow [{\displaystyle k_{\mbox{\scriptsize b}}}]{\hspace{6ex}}\hspace{1ex}\{AB\},
\end{equation}
since this implies that when the encounter partners separate, they immediately become part of the bulk solution. Reaction \ref{eq:Breakup} can only be a good approximation if the two encounter partners never meet again (as happens in the gas phase) or if the reaction probability during an encounter is very small (as happens in an activation-controlled reaction) so that becoming part of the bulk solution is the dominant fate of the molecules which separate from a non-reactive encounter.

A reaction scheme that includes re-encounters is depicted in Figure~\ref{fig:Complex02}.
\end{multicols}

\begin{figure}[!h] 
\setlength{\unitlength}{1mm}
\begin{picture}(100, 30)(-10,25)
\put(-10, 57){\line(1,0){165}}
\put(-10, 57){\line(0,1){2}}
\put(155, 57){\line(0,1){2}}
\put(-10, 12){\line(1,0){165}}
\put(-10, 12){\line(0,-1){2}}
\put(155, 12){\line(0,-1){2}}
\multiput(11.5,52)(1,0){45}
{\line(1,0){0.5}}
\multiput(11.5,24)(1,0){45}
{\line(1,0){0.5}}
\multiput(11.5,24)(0,1){28}
{\line(0,1){0.5}}
\multiput(56.5,24)(0,1){28}
{\line(0,1){0.5}}
\put(13.5, 39){\footnotesize(in the bulk)}
\put(15, 46){$\underbrace{
\hspace{1ex} A + B\hspace{1ex}  
}
\xrightarrow [\hspace{6ex}]{{\bf 1}} 
\big\{A\, B\big\}
\xrightarrow [\hspace{8ex}]{{\bf reaction}}
\big\{P\big\}
\xrightarrow [\hspace{6ex}]{{\bf 3}}
P
$}
\put(46, 37.5){\huge$\Downarrow$}
\put(50, 38){\huge$\Uparrow$}
\put(45, 32){$A \oplus B$}
\put(15, 32){$\overbrace{
\hspace{1ex} A + B\hspace{1ex} 
}
\xleftarrow [\hspace{6ex}]{{\bf 2}}
$}
\put(41, 28.5){\scriptsize non-contact}
\put(41.5, 25.5){\scriptsize neighbours}
\end{picture}
\caption{The principal events involved in a diffusion-mediated reaction. } 
\label{fig:Complex02}
\end{figure}
\begin{multicols}{2}
The reactants ($A$ and $B$), initially part of the bulk solution, become encounter partners (step {\bf 1}). Two fates are open to the encounter pair: reaction to form the product $P$, or breaking up into a pair of neighbouring molecules, denoted here as $A\oplus B$, which are too far apart to react together but close enough to be distinct from a statistically distributed pair of reactants. The two non-contact neighbours may become encounter partners once more, or may become part of the bulk solution (step {\bf 2}). The product molecule $P$ diffuses out of the cage where it was formed (step {\bf 3}), but these displacements are of no interest here, which means that no error arises if one replaces the two step reaction $\{AB\} \rightarrow\{P\} \rightarrow P$ by the single-step $\{AB\}\rightarrow P$. However, it is in general illicit to replace the events taking place within the dashed box in Figure~\ref{fig:Complex02} by the reaction $A+B\rightleftharpoons \{AB\}$. 

\vspace{0.5ex}
Noyes \cite{Noyes1954JCP} showed that the long-time value of the rate constant of a diffusion-mediated reaction can be expressed as
\begin{equation}\label{eq:Noyes01}
k= \alpha k_{\mbox{\scriptsize e}}\,\left [\frac{1-\beta}{1-\beta + \alpha\beta}\right ],
\end{equation}
where $\beta$ is the probability that two molecules separating from an encounter will re-encounter at least once more. Notice that the term within the square brackets reduces to unity if $\alpha \ll 1$ (activation-controlled reaction) or if $\beta$ is negligible (gas-phase reaction).

Workers who use scheme \ref{eq:ThreeArrows} and Eq.~\ref{eq:wrongrate} do so under the mistaken belief that both are consistent with the conclusions reached by C\&K and Noyes; such readers as are interested in finding the correct expression for $k$ should consult ref.~\cite{KRN1982JPC}.

\end{multicols}
}

\end{document}